\documentclass[aps,prl,twocolumn,showpacs,amsmath,amssymb,superscriptaddress]{revtex4}
\usepackage{graphicx}
\usepackage{amssymb}
\usepackage{natbib}
\usepackage{color}

\newcommand{\beq}{\begin{eqnarray}}
\newcommand{\eeq}{\end{eqnarray}}

\begin{document}
\title{Energy dependence of the spin excitation anisotropy in uniaxial-strained BaFe$_{1.9}$Ni$_{0.1}$As$_{2}$}
\author{Yu Song}
\affiliation{Department of Physics and Astronomy, Rice University, Houston, Texas 77005, USA
}
\author{Xingye Lu}
\affiliation{Department of Physics and Astronomy, Rice University, Houston, Texas 77005, USA
}
\affiliation{Beijing National Laboratory for Condensed Matter
Physics, Institute of Physics, Chinese Academy of Sciences, Beijing
100190, China}
\author{D. L. Abernathy}
\affiliation{Quantum Condensed Matter Division, Oak Ridge National Laboratory, Oak Ridge, Tennessee 37831, USA}
\author{David W. Tam}
\affiliation{Department of Physics and Astronomy, Rice University, Houston, Texas 77005, USA
}
\author{J. L. Niedziela}
\affiliation{Instrument and Source Division, Oak Ridge National Laboratory, Oak Ridge, Tennessee 37831, USA}
\author{Wei Tian}
\affiliation{Quantum Condensed Matter Division, Oak Ridge National Laboratory, Oak Ridge, Tennessee 37831, USA}
\author{Huiqian Luo}
\affiliation{Beijing National Laboratory for Condensed Matter
Physics, Institute of Physics, Chinese Academy of Sciences, Beijing
100190, China}
\author{Qimiao Si}
\affiliation{Department of Physics and Astronomy, Rice University, Houston, Texas 77005, USA
}
\author{Pengcheng Dai}
\email{pdai@rice.edu} 
\affiliation{Department of Physics and Astronomy, Rice University, Houston, Texas 77005, USA
}

\begin{abstract}
We use inelastic neutron scattering to study the temperature and energy dependence of the
spin excitation anisotropy in uniaxial-strained electron-doped iron pnictide BaFe$_{1.9}$Ni$_{0.1}$As$_2$ 
near optimal superconductivity ($T_c=20$ K). Our work has been motivated by the observation of
in-plane resistivity anisotropy in the paramagnetic tetragonal phase of electron-underdoped iron pnictides under uniaxial pressure, which has been attributed to a spin-driven Ising-nematic state or orbital ordering. 
Here we show that the spin excitation anisotropy, a signature of the spin-driven Ising-nematic phase, 
exists for energies below $\sim$60 meV 
in uniaxial-strained BaFe$_{1.9}$Ni$_{0.1}$As$_2$.  Since this energy scale is considerably larger 
than the energy splitting of the $d_{xz}$ and $d_{yz}$ bands of uniaxial-strained 
Ba(Fe$_{1-x}$Co$_x$)$_2$As$_2$ near optimal superconductivity, spin Ising-nematic correlations
is likely the driving force for the resistivity anisotropy and associated 
electronic nematic correlations.
\end{abstract}

\pacs{74.25.Ha, 74.70.-b, 78.70.Nx}

\maketitle

An electronic nematic phase, where the rotational symmetry of the system is spontaneously broken without 
breaking the translational symmetry of the underlying lattice \cite{fradkin}, has been
observed close to the superconducting
phase in iron pnictides \cite{fisher}.  In the undoped state, 
the parent compounds of iron pnictide superconductors such as BaFe$_2$As$_2$ 
exhibits a tetragonal-to-orthorhombic structural transition at $T_s$ that precedes
the onset of long-range collinear antiferromagnetic (AF) order below the ordering temperature $T_N$
 \cite{kamihara,stewart,dai,cruz,qhunag,mgkim}.
Upon electron-doping via partially replacing Fe by Co or Ni to form 
Ba(Fe$_{1-x}$Co$_x$)$_2$As$_2$ \cite{CLester2009,SNandi} 
or BaFe$_{2-x}$Ni$_x$As$_2$ \cite{hqluo,xylu13}, 
both $T_s$ and $T_N$ are suppressed with increasing 
doping leading to superconductivity [Fig. 1(a)].
A key signature of electronic nematicity 
has been 
the in-plane resistivity anisotropy found in Ba(Fe$_{1-x}$Co$_x$)$_2$As$_2$ under uniaxial pressure above the superconducting 
transition temperature $T_c$, stress-free $T_N$ and $T_s$ \cite{jhchu,matanatar,chu12}.  In particular, recent 
elastoresistance \cite{chu12,HHKuo2014,HHKuo2015} and elastic moduli \cite{Yoshizawa,bohmer14} 
measurements on Ba(Fe$_{1-x}$Co$_x$)$_2$As$_2$ reveal a divergence of the 
electronic nematic susceptibility, defined as the susceptibility of electronic anisotropy to anisotropic in-plane
strain, upon approaching $T_s$.  While these results indicate 
that the structural phase transition is driven by electronic  
degrees of freedom, it is still unclear whether it is due to the 
spin Ising-nematic state that breaks the in-plane four-fold rotational symmetry
of the underlying paramagnetic tetragonal lattice \cite{CCL,si,jphu,fernandes11,fernandes12,fernandes12a}, 
or arises from the orbital ordering of 
Fe  $d_{xz}$ and $d_{yz}$ orbitals among the five Fe $3d$ orbitals \cite{cclee,kruger,lv,ccchen,valenzeula}.

Experimentally, inelastic neutron scattering (INS) experiments on BaFe$_{2-x}$Ni$_x$As$_2$ ($x=0,0.085,0.12$) under uniaxial pressure indicate that spin excitations at energies below 16 meV change from four-fold symmetric to two-fold
symmetric in the tetragonal phase at temperatures approximately 
corresponding to the onset of the in-plane resistivity anisotropy, thus suggesting 
that the spin Ising-nematic correlations is associated with the resistivity anisotropy \cite{xylu14}.
On the other hand, X-ray linear dichroism (XLD) \cite{YKKim} and 
angle resolved photoemission spectroscopy (ARPES) \cite{myi,yzhang}
experiments indicate the tendency towards
orbital ordering in the tetragonal phase of Ba(Fe$_{1-x}$Co$_x$)$_2$As$_2$ under uniaxial pressure. 
In particular, an in-plane electronic anisotropy, characterized by a
$\sim$60 meV energy splitting of two orthogonal bands with dominant
$d_{xz}$ (${\bf Q}_2$) and $d_{yz}$ (${\bf Q}_1$) character in the AF ordered 
orthorhombic state of undoped BaFe$_2$As$_2$ 
and underdoped Ba(Fe$_{1-x}$Co$_x$)$_2$As$_2$ [Fig. 1(c), 1(d)], is observed to 
develop above the stress-free $T_N$ and $T_s$  
similar to the resistivity anisotropy [Fig. 1(e)] \cite{myi}.  
Furthermore, the uniaxial pressure necessary to detwin  Ba(Fe$_{1-x}$Co$_x$)$_2$As$_2$ or BaFe$_{2-x}$Ni$_x$As$_2$ iron pnictides can also
affect their transport properties \cite{man}, and magnetic \cite{Dhital12,Dhital14} and structural \cite{xylu15} 
phase transitions. Therefore, it remains unclear if the electronic nematic phase is due to the spin 
Ising-nematic state \cite{CCL,si,jphu,fernandes11,fernandes12,fernandes12a}, 
orbital ordering \cite{cclee,kruger,lv,ccchen,valenzeula}, or applied uniaxial strain 
via enhanced spin or orbital nematic susceptibility.

One way to reveal whether the spin Ising-nematic state is associated with orbital ordering or not 
 is to determine the energy dependence of the 
spin excitation anisotropy and its electron doping dependence. 
By determining the energy and temperature dependence of the spin excitation anisotropy, one can compare 
the outcome with  temperature and electron-doping 
dependence of the energy splitting of the $d_{xz}$ and $d_{yz}$ bands 
in Ba(Fe$_{1-x}$Co$_x$)$_2$As$_2$ \cite{myi}, and therefore establish
whether and how the spin Ising-nematic correlations are associated with orbital ordering \cite{Kovacic}.

\begin{figure}[t]
\includegraphics[scale=.75]{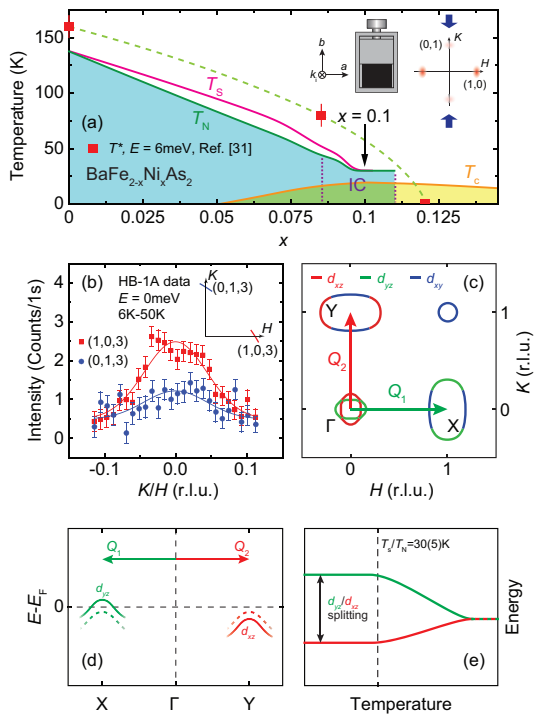}
\caption{
(Color online) (a) The phase diagram of BaFe$_{2-x}$Ni$_{x}$As$_2$. In BaFe$_{1.9}$Ni$_{0.1}$As$_2$ superconductivity coexists with incommensurate (IC) short-range magnetic order \cite{hqluo}. The mechanical clamp used and the magnetic excitations under uniaxial pressure along \textit{b} axis are schematically shown in the inset at the top-right. The red squares and dashed line mark $T^\ast$, a crossover temperature at which intensity of low-energy magnetic excitations at $(1,0)$ and $(0,1)$ in BaFe$_{2-x}$Ni$_{x}$As$_2$ under uniaxial pressure merge \cite{xylu14}. (b) Rocking scans of the elastic magnetic peak at 6 K obtained on HB-1A, background measured at 50 K has been subtracted. The inset shows the rocking scans projected into the $[H,K,0]$ plane. (c) Schematic Fermi surface of BaFe$_{1.9}$Ni$_{0.1}$As$_2$ in the paramagnetic state, the arrows mark nesting wave vectors ${\bf Q}_{1}=(1,0)$ and ${\bf Q}_{2}=(0,1)$. Fermi surfaces originating from different orbitals are shown in different colors. (d) Schematic splitting of $d_{yz}$ and $d_{xz}$ bands at $X$ and $Y$ in Ba(Fe$_{1-x}$Co$_x$)$_2$As$_2$, as found by ARPES \cite{myi}. At higher temperatures, the two bands have the same energy 
(dashed lines) but as temperature is lowered $d_{yz}$ band moves up in energy whereas $d_{xz}$ move down. (e) Schematic temperature dependence of the orbital splitting in (d), under uniaxial pressure the splitting persists to above the stress-free $T_N$ and $T_s$.
 }
 \end{figure}

In this paper, we report INS studies of temperature and energy evolution of the spin excitation anisotropy 
in superconducting BaFe$_{1.9}$Ni$_{0.1}$As$_2$ ($T_c=20$ K, $T_N\approx T_s\approx 30\pm 5$ K) 
detwinned under uniaxial pressure \cite{xylu13,msliu,xylu14a,daLuz,Simayi}. We chose 
to study BaFe$_{1.9}$Ni$_{0.1}$As$_2$ because 
ARPES measurements on Ba(Fe$_{1-x}$Co$_{x}$)$_2$As$_2$
samples reveal vanishing energy splitting of the $d_{xz}$ and $d_{yz}$ bands ($\sim$20 meV) and orbital ordering 
approaching optimal doping \cite{myi}.
Using time-of-flight neutron spectroscopy, we show that the spin excitation anisotropy 
in BaFe$_{1.9}$Ni$_{0.1}$As$_2$ in the low-temperature superconducting state 
decreases with increasing energy, and vanishes for energies above $\sim$60 meV (Fig. 2).
This anisotropy energy scale is remarkably similar to the energy splitting ($\sim$65 meV)
of the $d_{xz}$ and $d_{yz}$ bands seen by ARPES in
the undoped and electron underdoped Ba(Fe$_{1-x}$Co$_{x}$)$_2$As$_2$ iron pnictides [Fig. 1(e)] \cite{myi}. 
Upon warming to high temperatures, the spin 
excitation anisotropy at $E=4.5\pm 0.5$ meV decreases smoothly with increasing temperature 
showing no anomaly across $T_c$, stress-free $T_N$ and $T_s$, and 
vanishes around a crossover temperature $T^\ast$, where resistivity anisotropy vanishes (Fig. 3) \cite{xylu14}.
The energy dependence of the spin excitation anisotropy, however, is weakly temperature 
dependent from 5 K ($\ll T_c$) to 35 K ($>T_N,T_s$), and persists below 60 meV. 
Since the energy splitting of the $d_{xz}$ and $d_{yz}$ orbitals decreases with increasing 
electron-doping for
Ba(Fe$_{1-x}$Co$_{x}$)$_2$As$_2$ and diminishes rapidly above $T_N$  \cite{myi}, our observation of the 
large energy ($\sim$60 meV) spin excitation anisotropy in the uniaxial strained paramagnetic state of a 
nearly optimally electron-doped BaFe$_{1.9}$Ni$_{0.1}$As$_2$ is larger than the energy splitting of optimally
doped iron pnictides above $T_N$, thus suggesting that 
the spin Ising-nematic state may be the driving force for the  
electronic nematicity in iron pnictides \cite{CCL,si,jphu,fernandes11,fernandes12,fernandes12a}.

\begin{figure}[t]
\includegraphics[scale=.5]{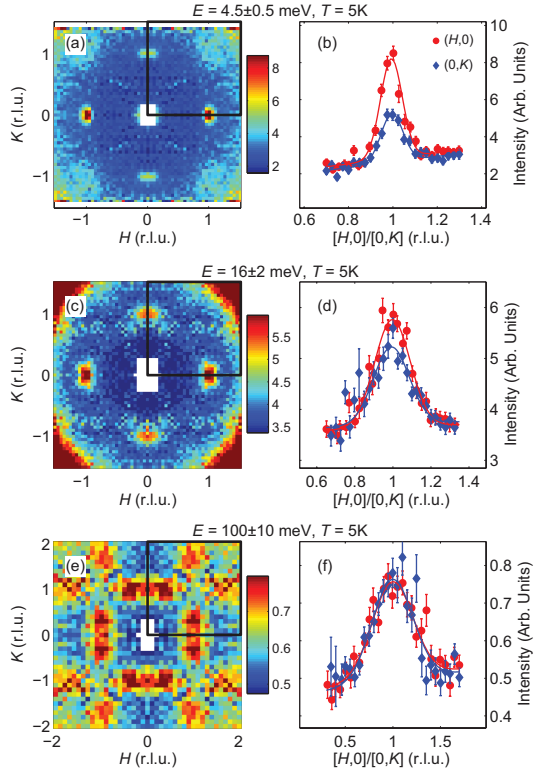}
\caption{ 
(Color online) Constant-energy slices symmetrized along $H$ and $K$ axes at $T=5$ K for energy transfers (a) $E=4.5 \pm 0.5$ meV ($E_i=30$ meV), (c) $E=16 \pm 2$ meV ($E_i=80$ meV) and (e) $E=100 \pm 10$ meV ($E_i=250$ meV). The black boxes indicate regions that contain non-duplicate data due to symmetrizing. Longitudinal cuts along $[H,0]$ (red circles) and $[0,K]$ (blue diamonds) for energy transfers in (a), (c) and (e) are respectively shown in (b), (d) and (f). The solid lines are fits using Gaussian functions and linear backgrounds. $[H,0]$/$[K,0]$ scans are are obtained by binning $K$/$H$ in the range (b) $[-0.15,0.15]$, (d) $[-0.175,0.175]$ and (f)$[-0.3,0.3]$ and folding along $(K,0)$/$(H,0)$. 
 }
 \end{figure}

\begin{figure}[t]
\includegraphics[scale=.5]{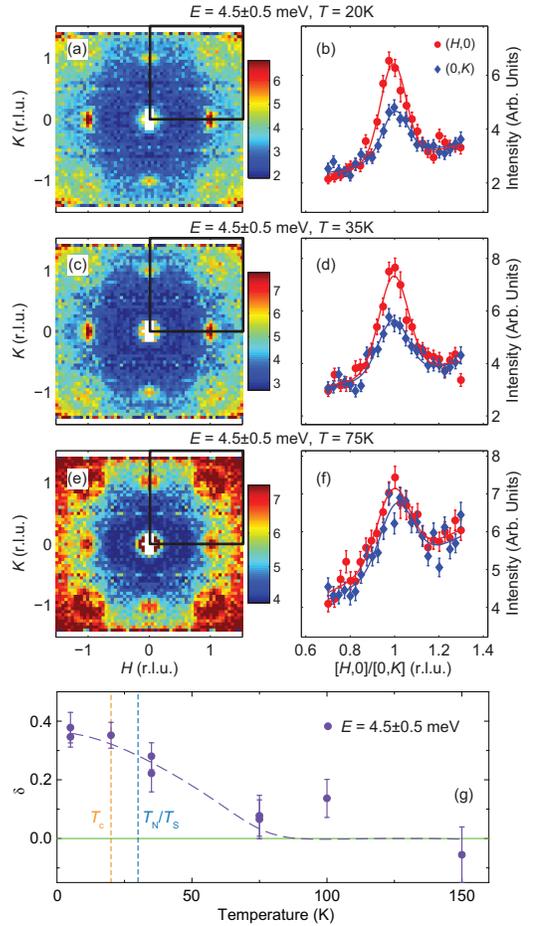}
\caption{
(Color online) Constant-energy slices symmetrized along $H$ and $K$ axes for $E=4.5 \pm 0.5$ meV ($E_i=30$meV) at (a) 20 K, (c) 35 K and (e) 75 K. Corresponding longitudinal cuts along $[H,0]$ (red circles) and $[0,K]$ (blue diamonds) are respectively shown in (b), (d) and (f). $[H,0]$/$[K,0]$ scans are are obtained by binning $K$/$H$ in the range $[-0.15,0.15]$. (g) Temperature dependence of the anisotropy $\delta=(I_{10}-I_{01})/(I_{10}+I_{01})$ for $E=4.5 \pm 0.5$meV. The purple dashed line is a guide to the eye. $T_c$ and stress-free $T_N$/$T_s$ are marked by vertical dashed lines.
 }
\end{figure}

Our neutron scattering experiments were carried out at the Wide Angular-Range Chopper Spectrometer (ARCS) 
at the Spallation Neutron Source  
and HB-1A triple-axis spectrometer at the High-Flux Isotope Reactor, Oak Ridge National Laboratory. 
The BaFe$_{1.9}$Ni$_{0.1}$As$_2$ single crystals \cite{msliu,xylu14a} are cut along the $a$, $b$ axes 
and each cut sample is loaded into an individual mechanical clamp with applied uniaxial pressure \cite{supplementary}. 9 crystals with a total mass ~6.5 grams were co-aligned. Elastic neutron scattering measurements were carried out on HB-1A to determine the detwinning ratio in the orthorhombic phase. 
The momentum transfer ${\bf Q}$ in three-dimensional reciprocal space in \AA$^{-1}$ is defined 
as $\textbf{Q}=H\textbf{a}^\ast+K\textbf{b}^\ast+L\textbf{c}^\ast$, where $H$, $K$, and $L$ are Miller indices and 
${\bf a}^\ast=\hat{{\bf a}}2\pi/a$, ${\bf b}^\ast=\hat{{\bf b}}2\pi/b$, ${\bf c}^\ast=\hat{{\bf c}}2\pi/c$ with  
$a\approx b= 5.564$ \AA, and $c=12.77$ \AA.
In the AF ordered state of a fully detwinned sample, the AF Bragg peaks should occur at $(\pm 1,0,L)$ ($L=1,3,5,\cdots$) 
positions in reciprocal space \cite{qhunag}. 
For elastic neutron scattering measurements on HB-1A, the samples are aligned in 
the scattering plane spanned by the wave vectors $(1,0,3)$ and $(0,1,3)$ with $E_i = 14.6$ meV.
Figure 1(b) shows elastic scans through the 
$(1,0,3)$ and $(0,1,3)$ positions
to obtain the ratio ($R=I_{10}/I_{01}$) of magnetic intensities.
Two Gaussians with linear backgrounds having the same widths and backgrounds were fit to scans as solid lines [Fig. 1(b)]. 
Anisotropy of intensities between ${\bf Q}_1=(1,0)$ and ${\bf Q}_2=(0,1)$ is then 
obtained through $\delta=(I_{10}-I_{01})/(I_{10}+I_{01})=(R-1)/(R+1)\approx 0.5$.
In a fully detwinned sample, one would expect $\delta \rightarrow 1$, while in a completely twinned sample $\delta \rightarrow 0$. In a partially detwinned sample with volume fraction of $x$ corresponding to
magnetic order at $(1,0)$, the actual observed spin excitation intensities 
at $(1,0)$ and $(0,1)$ should respectively be $I_{10}=x \widetilde{I}_{10}+(1-x)\widetilde{I}_{01}$ and 
$I_{01}=x \widetilde{I}_{01}+(1-x) \widetilde{I}_{10}$, with $\widetilde{I}_{10}$ and $\widetilde{I}_{01}$ being the spin excitation intensity at $(1,0)$ and $(1,0)$ in a fully detwinned sample.  Therefore, for a partially
detwinned sample, one has $\delta=(I_{10}-I_{01})/(I_{10}+I_{01})=(2x-1)\widetilde{\delta}$ with $\widetilde{\delta}=(\widetilde{I}_{10}-\widetilde{I}_{01})/(\widetilde{I}_{10}+\widetilde{I}_{01})$.  This means regardless of
the detwinning ratio, $\delta$ is directly proportional to $\widetilde{\delta}$ and the energy/temperature dependence of experimentally obtained $\delta$ display the intrinsic behavior of $\widetilde{\delta}$ even for a partially detwinned sample. For the ARCS experiment, incident beam is directed along \textit{c} axis of the 
samples and incident energies of $E_i=30$, 80, 150 and 250 meV were used.  The observed magnetic scattering $I_{10}$ and
$I_{01}$ are related to the imaginary part of the dynamic susceptibility 
$\chi^{\prime\prime}_{10}$ and $\chi^{\prime\prime}_{01}$, respectively, 
via the Bose factor \cite{dai}.

\begin{figure}[t]
\includegraphics[scale=.83]{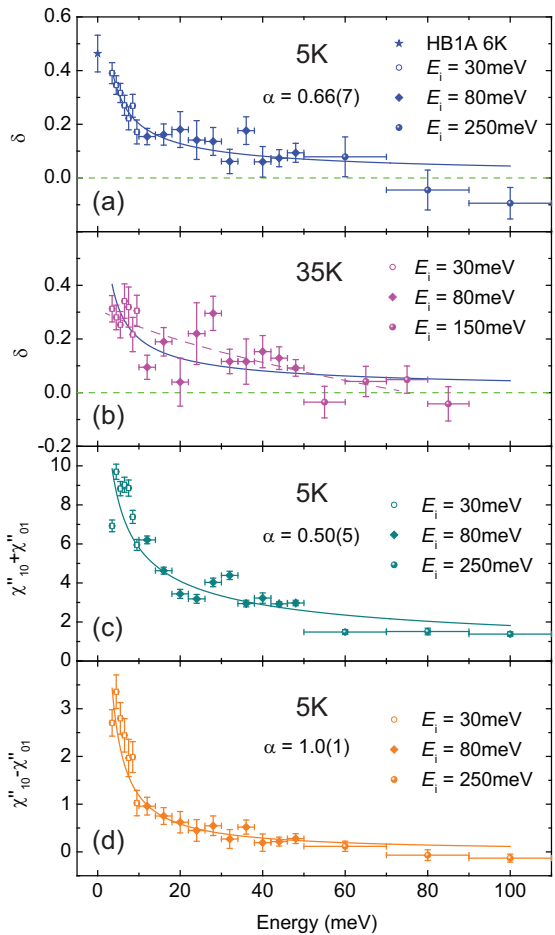}
\caption{
(Color online) Energy dependence of anisotropy between ${\bf Q}_1$ and ${\bf Q}_2$ defined as $\delta=(I_{10}-I_{01})/(I_{10}+I_{01})$ for (a) 5 K and (b) 35 K. (c) Energy dependence of $\chi^{\prime\prime}_{10}+\chi^{\prime\prime}_{01}$ at 5K, $\chi^{\prime\prime}_{10}$ and $\chi^{\prime\prime}_{01}$ are dynamic susceptibilities at ${\bf Q}_1=(1,0)$ and ${\bf Q}_2=(0,1)$. (d) 
$\chi^{\prime\prime}_{10}-\chi^{\prime\prime}_{01}$. 
Data obtained on HB-1A is collected at 6 K, and is plotted together with ARCS data 
using incident energies $E_i$ = 30, 80, 150 and 250meV. 
 }
\end{figure} 

Figures 2(a), 2(c), and 2(e) show constant-energy slices of spin excitations in
BaFe$_{1.9}$Ni$_{0.1}$As$_2$ in the $(H,K)$ plane
 at 5 K for energy transfers $E=4.5\pm 0.5$, $16\pm 2$, and $100\pm 10$ meV, respectively.
For $E=4.5\pm0.5$ meV, the scattering intensity at ${\bf Q}_1=(\pm 1,0)$ is much stronger 
than at ${\bf Q}_2=(0,\pm 1)$ [Fig. 2(a)] \cite{xylu14}. 
Figure 2(b) compares constant-energy cuts
along the $[H,0]$ and $[0,K]$ directions, confirming the stronger intensity at $(1,0)$.  On increasing
the energy to $E=16\pm 2$ meV, the intensity difference between ${\bf Q}_1=(\pm 1,0)$ and 
${\bf Q}_2=(0,\pm 1)$ becomes smaller [Fig. 2(c)], as revealed in constant-energy cuts of Fig. 2(d). 
At an energy transfer of $E=100\pm 10$ meV, 
the scattering becomes isotropic, and no discernible difference can be seen at ${\bf Q}_1=(\pm 1,0)$ and ${\bf Q}_2=(0,\pm 1)$ [Fig. 2(e)].
This is confirmed by constant-energy cuts along the $[H,0]$ and $[0,K]$ directions [Fig. 2(f)].

Figure 3 shows  
constant-energy slices of spin excitations with $E=4.5\pm 0.5$ meV on warming from $T=20$ K to 75 K.
At $T=20$ K ($T_s\geq T_N>T>T_c$), the spin excitation anisotropy 
shown in Fig. 3(a) and Fig. 3(b) is similar to $T=5$ K [Fig. 2(a) and 2(b)].
On warming to $T=35$ K ($T>T_s\geq T_N>T_c$) corresponding to the tetragonal state in stress-free samples, 
clear differences in spin excitation intensity between ${\bf Q}_1=(\pm 1,0)$ and 
${\bf Q}_2=(0,\pm 1)$ can be still seen [Fig. 3(c) and 3(d)].  The differences 
between these two wave vectors essentially disappear at $T=75$ K, a temperature well 
above the strain-free $T_s$ and $T_N$ [Fig. 3(e) and 3(f)].  The spin excitation 
anisotropy $\delta$ decreases smoothly with increasing temperature and vanishes around 80 K [Fig. 3(g)], 
similar to the resistivity anisotropy \cite{xylu14}. 

To quantitatively determine the energy and temperature dependence of spin excitation anisotropy, we systematically made constant-energy slices and cuts along $[H,0]$ and $[0,K]$ at various energies similar to 
Figs. 2 and 3. Based on the cuts, we 
can estimate the energy dependence of the spin excitation anisotropy $\delta$ \cite{supplementary}.
Figure 4(a) shows 
that the spin excitation anisotropy ($\delta$) decreases with increasing energy and
vanishes for energy transfers above $\sim$60 meV at $T=5$ K ($\ll T_c, T_N, T_s$). On warming to 35 K, 
a temperature above $T_c$, $T_N$, and $T_s$, 
the energy of the spin excitation anisotropy still persists to about $\sim$60 meV, similar to 5 K [Fig. 4(b)]. 
  
We are now in a position to compare and contrast our results with the orbital ordering 
tendencies indicated by the ARPES measurements  \cite{myi}.
The energy splitting of the $d_{xz}$ and $d_{yz}$ bands in undoped 
and underdoped Ba(Fe$_{1-x}$Co$_{x}$)$_2$As$_2$ is also about $\sim$60 meV, 
and is likewise weakly temperature dependent below $T_s$ [Fig. 1(e)] \cite{myi}. Upon increasing the doping level to near optimal superconductivity, the ARPES-measured orbital splitting energy 
in electron-doped iron pnictides decreases to $\sim$20 meV and vanishes very rapidly above $T_N,T_s$ \cite{myi}.
Since the ARPES-measured orbital splitting energy \cite{myi} and neutron scattering measured spin excitation
anisotropy \cite{xylu14} in the paramagnetic state may be uniaxial strain dependent \cite{man}, it would be more constructive to compare the doping dependence of the spin excitation anisotropy in the uniaxial strained 
paramagnetic state with those of APRES measurements. 
For BaFe$_2$As$_2$, our unpublished results suggest spin excitation anisotropy persists to about $60$ meV at 145 K (just above
$T_N,T_s$ of 140 K) \cite{dai}.  For BaFe$_{1.9}$Ni$_{0.1}$As$_2$, $\delta$ is also nonzero 
below $\sim$60 meV both below and above
$T_N,T_s$ [Fig. 4(a) and 4(b)].  This means that spin excitations anisotropy is weakly doping dependent and has a larger anisotropy energy scale than that of the ARPES-measured orbital splitting energy, suggesting that 
it is likely the spin channel, instead of the orbital sector, that drives the Ising-nematic correlations.

To further analyze the energy dependence of the spin correlations,
we show in Figures 4(c) and 4(d) the energy dependence of the sum, $\chi^{\prime\prime}_{10}+\chi^{\prime\prime}_{01}$,
and difference, $\chi^{\prime\prime}_{10}-\chi^{\prime\prime}_{01}$, of the dynamic susceptibilities at the two wave vectors (For the measured energy and temperature range, $\chi^{\prime\prime}({\bf Q},\omega)$ is directly proportional to the measured neutron scattering intensity assuming the magnetism is essentially two-dimensional and after correcting for the magnetic form factor), respectively. It is seen that both quantities 
increase as energy is decreased. Within the measured energy range, both the sum and difference 
can be fit with a power-law dependence on the energy, $\sim 1/E^{\alpha}$, with exponents $\alpha$ being 
$0.50(5)$ and $1.0(1)$ respectively. The ratio, $\delta$, can also be fitted with a power-law divergence, 
although this divergence must be truncated at frequencies below the measured low-frequency limit,
 because $\delta$ must be bound by $1$.

It is instructive to contrast the spin nematic scenario with an alternative picture based on orbital ordering.
 Since the electron-doping evolution of the 
low-energy spin excitations in BaFe$_{2-x}$Ni$_x$As$_2$
is consistent with quasiparticle excitations between the hole Fermi surfaces near
$\Gamma$ and electron Fermi surfaces at ${\bf Q}_1=(1,0)$ (${\bf Q}_2=(0,1)$) [Fig. 1(c)] \cite{hqluo12}, 
an energy splitting of the $d_{xz}$ and $d_{yz}$ bands at these two wave vectors 
should result in spin excitation anisotropy as seen by INS \cite{Kovacic}.
However, this picture would require that the tendency towards the orbital ordering is stronger
than the spin-excitation anisotropy, which is opposite to our results near 
 the optimal electron doping. Nevertheless, since spin and orbital degrees of freedom 
in iron pnictides are generally coupled, 
it may not be experimentally possible to conclusively determine if 
spin or orbital degrees of freedom is the driving force for the enhanced 
nematic susceptibility.

In summary, we have discovered that the four-fold symmetric to two-fold symmetric transition
of spin excitations in BaFe$_{2-x}$Ni$_{x}$As$_2$ under uniaxial pressure is energy dependent
and occurs for energy transfers below
about 60 meV in near optimally electron-doped iron pnictides.  
Since orbital splitting becomes vanishingly small for optimally
electron-doped iron pnictides in the paramagnetic state of uniaxial strained sample,
 our results would suggest that the spin excitation anisotropy or spin Ising-nematic correlations
  is the driving force for the  
electronic nematic correlations in iron pnictides.

The neutron work at Rice is supported by the
U.S. NSF-DMR-1362219 and DMR-1436006 (P.D.).  This work is also supported by 
the Robert A. Welch Foundation Grant Nos. C-1839 (P.D.) and C-1411 (Q.S.).
Q.S. is supported by the U.S. NSF-DMR-1309531. The neutron work at ORNL's 
HFIR and SNS was sponsored by the Scientific User Facilities Division, Office of Basic Energy Sciences, US department of Energy.

\end{document}